# A Prediction Model for System Testing Defects using Regression Analysis

[1]Muhammad Dhiauddin Mohamed Suffian, [2]Suhaimi Ibrahim
[1]*Faculty of Computer Science & Information System,
Universiti Teknologi Malaysia*
[2]*Advanced Informatics School, Universiti Teknologi Malaysia*
Email: [1]mdhiauddin2@live.utm.my, [2]suhaimiibrahim@utm.my

*Abstract.* This research describes the initial effort of building a prediction model for defects in system testing carried out by an independent testing team. The motivation to have such defect prediction model is to serve as early quality indicator of the software entering system testing and assist the testing team to manage and control test execution activities. Metrics collected from prior phases to system testing are identified and analyzed to determine the potential predictors for building the model. The selected metrics are then put into regression analysis to generate several mathematical equations. Mathematical equation that has p-value of less than 0.05 with R-squared and R-squared (adjusted) more than 90% is selected as the desired prediction model for system testing defects. This model is verified using new projects to confirm that it is fit for actual implementation.

*\*Corresponding address:*
Muhammad Dhiauddin Mohamed Suffian,
*Faculty of Computer Science & Information System,
Universiti Teknologi Malaysia,
Malaysia*
Email: mdhiauddin2@live.utm.my

## 1. Introduction

As independent testing team, it is important to plan and manage the test execution activities in order to meet the tight deadline for releasing the software to end-users. Since the aim of test execution is to discover as many defects as possible, testing team is usually put into burden to ensure all defects are found and fixed by the developers within the system testing phase. Additional number of days has to be added to the timeline to accommodate testing team in completing their test with the hope that all defects have been found and fixed. On the other hand, the stakeholders would also ask the testing team on the forecasted defects in the software so that they could decide whether the software is feasible and fit for release. This is due to the nature that system testing is the last gate before the software is made visible to end-users, thus as the custodian of executing system testing, the independent testing team has to take responsibility to ensure software to be released is of high quality.

Therefore, the ability to predict how many defects that can be found at the start of system testing shall be a good way to tackle this issue. This becomes the reason for conducting this study. Besides serving as a target on how many defects to capture in system testing, defect prediction can also become an early quality indicator for any software entering the testing phase. Testing team can use the predicted defects to plan, manage and control test execution activities. This could be in the form aligning the test execution time and number of test engineers assigned to particular testing project. Having defect prediction as part of the testing process allows testing team to strengthen their test strategies by adding more exploratory testing and user experience testing to ensure known defects are not escaped and re-introduced to end-users. Test engineers would be able to have better root cause analysis of the defects found. In the long run, testing can achieve what is called as zero-known post release defects for that particular software.

This research is driven towards achieving several objectives as follows:
- To analyse existing metrics, techniques and approaches used in building prediction model for system testing defects
- To analyse the metrics in prior phases to testing that could be used for predicting defects in system testing
- To formulate prediction model for system testing defects using statistical approach based on metrics in prior phases to system testing
- To evaluate the accuracy of proposed prediction model based on acceptable criteria for final selection of defect prediction model for system testing

## 2. Related Works

Software defect prediction is not a new thing in software engineering domain. Various related studies and approaches have been conducted to come out with the right defect prediction model. Understanding what defect really means is important so that the term is not confused with error, mistake or failure. In the event when the software or system fails to perform its desired function, it means that defects have taken place [1]. Defect is also observed as the deviation from its specification [2] as well as any imperfection related to software itself and its related work product [3]. Therefore, anything that is not according to specification for software and its work product is referred as defect. Knowing what defect means is not enough since in building the prediction model for defects, it is essential to know how defects are introduced as part of verification and validation (V&V) activities [3].

Predicting defects is the proactive process of characterizing many types of defects found in software's content, design and codes in producing high quality product [4]. [2] presented that size and complexity metrics are among the earlier approaches to defect prediction. Lines of code (LOC) and McCabe's cyclomatic complexity were used to predict defects in software. Two of the equations presented were as follows:

$$\text{Defect} = 4.86 + 0.018 \text{ Lines of Code}$$
$$\text{Defect} = 4.2 + 0.0015 \text{ Lines of Code}^{4/3}$$

[2] also added that defects can also be predicted based on defect per life cycle prediction and defect found per testing approach. Apart from that, [3] categorized the techniques for predicting defects into three areas: project management, work product assessment and process improvement such as defect discovery profile, fault proneness evaluation, orthogonal defect classification and empirical defect prediction.

Rayleigh model was also used to predict defect density for different phases of project life cycle [5]. [6] constructed the model to predict defects using product and project metrics collected from design review, code testing, code peer review as well as product release usage and defect validation. Linear regression was applied to these metrics via product metrics only, project metrics only and both. As the result, linear regression using both product and project metrics provided better correlation between defects and the predictors. At the same time, it demonstrated the feasibility of using regression analysis to build defect prediction model. An approach was carried out to predict defects using mathematical distributions that serve as quality prediction model [7]. Further investigation was performed to predict which part of large multi-release industrial software system contains the highest defects in the next release. Result of the investigation stated that important factor for the prediction and its impact to the model quality is development information which focuses on three metrics: number of developers who modified the file during the prior release; the number of new developers who modified the file during the prior release; and the cumulative number of distinct developers who modified the file during all releases through the prior release [8].

There was also study to investigate on how to defect fault-proneness in the source code of the open source Web and e-mail suite called Mozilla. It used object-oriented metrics proposed by Chidamber and Kemerer to conduct the investigation [9]. On the other hand, [10] proposed several inputs to simulate the system test phase, in which those inputs could be considered as potential predictors to build defect prediction model. Another approach to defect prediction was based on simple Bayesian Network in a form of Defect Type Model (DTM) that predicts defects based on severity minor, major and minor [11]. Multivariate linear regression was used by [12] to come out with defect inflow prediction for large software projects either short-term defect inflow prediction or long-term defect inflow prediction.

[13] applied statistical approach in Six Sigma methodology to predict defect density. In this case, statistical method was used against the function point as the base metrics to predict defect density before releasing software to production. Defect prediction can also be observed from different perspective which is by predicting remaining total number of defects while the testing activities are still on-going [14], which is called as defect decay model. This model depends on on-going test execution data instead of historical data. [15] presented the case studies on building and using the defect prediction model in assisting their organization to assess testing effectiveness and predict the quantity of post release defects and enables quantitative decision about production go-live readiness. Their model was mainly focused on predicting defects in acceptance test or production which involves estimating total potential defects based on defined detailed requirements, applying defect removal efficiency and finally estimating the defects per phase as well as post release defects. The model implementation demonstrated a 1% defect removal efficiency improvement which equals to $20,000. However, if historical data is not available, having defect prediction would be difficult. Sample-based defect prediction was proposed to overcome this difficulty by using a small sample of modules to construct cost-effective defect prediction models for large scale systems, in which CoForest, a semi-supervised learning method was applied [16]. Knowing that defect prediction could optimize testing resources allocation, [17] conducted a feasibility study on predicting defects of cross-project when historical data is not in place. The results demonstrated that training data is very important for machine learning based defect prediction provided that the data is carefully selected from the projects. As an evidence, defects in 18 out of 34 cases from the study were predicted at a Recall greater than 70% and a Precision greater than 50%.

In building defect prediction model, it is essential to couple it with the mechanism to measure its success. [18] proposed to measure the percent of faults found in the identified files as one of the ways to assess the effectiveness of the prediction models. Besides that, the model is said to be a good model if it can assist in planning the resource for maintenance as well as in the area of software insurance towards coming out with software insurance system [19]. However, it is hard to find an established benchmark specific for defect prediction. An effort was taken by providing an extensive comparison of well-known bug prediction approaches, together with novel approaches using publicly available dataset consisting of several software systems [20]. The findings showed that there is still a problem with regard to external validity in defect prediction. It requires larger shared data set towards having a significant benchmark of defect prediction.

Looking at these related works, there is an opportunity to explore in the area of defect prediction model specific for system testing by applying existing proven technique, which is regression analysis to develop such model.

## 3. Data Requirement and Collection

An independent testing team from a research organization has been identified as the source of data. Therefore, the data collected for this research is subjected to metrics being captured, tracked and analyzed in their day-to-day software development activities. Looking at this constraint, the data may not be as extensive as those presented in other related works.

The data collected is based on the software development using V-shape development model which is used by the organization in developing and releasing the software. Implementation of V-shaped development model demonstrates rigor verification and validation activities that involvers proper collection of metrics in requirement and design review, test plan and test cases review, code inspection and unit testing, as well as the system testing itself. Looking at the requirement of this research which is to formulate a model to predict defects in system testing, therefore the factors should come from metrics collected in phases prior to system testing. As for the type of software which data is collected, it comprises of web-based and component-based developed in Java, PHP or Hypertext Preprocessor and .NET. The factors are categorized to several areas as outlined in Table 1 below:

Table 1. Possible factors to defect prediction

| Area | Factors |
|---|---|
| Software complexity | Number of requirement pages |
| | Number of design pages |
| | Type of programming language |
| | Code size |
| Knowledge | Developer knowledge |
| | Tester knowledge |
| Test process | Test case coverage |
| | Total test cases |
| | Test automation rate |
| | Test case execution productivity |
| | Total effort in test case design |
| | Total effort in phases prior to system testing |
| Errors | Requirement error |
| | Design error |
| | Code error |
| | Test plan error |
| | Test cases error |
| Fault | Requirement fault |
| | Design fault |
| | Code fault |
| | Integration fault |
| | Test cases fault |
| Defect | Severity of defect |
| | Type/category of defect |
| | Validity of defect |
| | Total defects logged |
| Type of software | Component-based |
| | Web-based |

Error in Table 1 means defect found within the same phase, defect means those captured within system testing while fault is the result of error plus defect.

After brainstorming activities, following factors were considered for further analysis in which the metrics were collected and analyzed:
- Number of requirement pages
- Number of design pages
- Code size in a form of lines of code
- Total test cases
- Total effort in test case design
- Total effort in phases prior to system testing
- Requirement error
- Design error
- Code error
- Test cases error
- Total defects logged in a form of all defects and functional defects

The data set collected is presented below in Table 2 after went through filtering process:

Table 2. Data set for regression analysis

| | Req. Error | Design Error | Coding Error | KLOC | Req. Page | Design Page | Total Test Cases | Test Cases Error | Total Effort | Test Design Effort | Functional Defects | All Defects |
|---|---|---|---|---|---|---|---|---|---|---|---|---|
| Project A | 5 | 22 | 12 | 28.8 | 81 | 121 | 224 | 34 | 16.79 | 15.20 | 19 | 19 |
| Project B | 0 | 0 | 1 | 6.8 | 171 | 14 | 17 | 6 | 45.69 | 40.91 | 1 | 1 |
| Project C | 9 | 10 | 14 | 5.4 | 23 | 42 | 24 | 6 | 13.44 | 13.44 | 4 | 4 |
| Project D | 7 | 12 | 2 | 1.1 | 23 | 42 | 25 | 9 | 4.90 | 4.90 | 0 | 0 |
| Project E | 11 | 29 | 3 | 1.2 | 23 | 54 | 28 | 12 | 4.72 | 4.59 | 3 | 3 |
| Project F | 0 | 2 | 7 | 6.8 | 20 | 70 | 66 | 7 | 32.69 | 16.00 | 16 | 27 |
| Project G | 3 | 25 | 11 | 4 | 38 | 131 | 149 | 0 | 64.00 | 53.50 | 3 | 3 |
| Project H | 4 | 9 | 2 | 0.2 | 26 | 26 | 24 | 0 | 5.63 | 5.63 | 0 | 0 |
| Project K | 17 | 0 | 3 | 1.4 | 15 | 28 | 13 | 4 | 9.13 | 7.88 | 1 | 1 |
| Project N | 61 | 34 | 24 | 36 | 57 | 156 | 306 | 16 | 89.42 | 76.16 | 25 | 28 |
| Project O | 32 | 16 | 19 | 12.3 | 162 | 384 | 142 | 0 | 7.00 | 7.00 | 12 | 12 |
| Project P | 0 | 2 | 3 | 3.8 | 35 | 33 | 40 | 3 | 8.86 | 8.86 | 6 | 6 |
| Project Q | 15 | 18 | 10 | 26.1 | 88 | 211 | 151 | 22 | 30.99 | 28.61 | 39 | 57 |
| Project R | 0 | 4 | 0 | 24.2 | 102 | 11 | 157 | 0 | 41.13 | 28.13 | 20 | 33 |

## 4. Findings and Discussion

This solution on using regression analysis in building prediction model for system testing is selected as the suitable approach since multiple factors affecting the discovery of defects in system testing can be analyzed and the relationship between these factors can be observed. This means significant factors that have impact to defects in system testing can be seen and identified clearly. As regression analysis is selected as the technique to construct the defect prediction model for system testing, three criteria are used to evaluate the outcome of the analysis:
- P-value – It determines the significance of the predictors to the discovery of functional defects. P-value must be less than 0.05
- R-Squared (R-Sq) –It is the amount of variation explained by the regression equation which is used to predict future outcomes on the basis of other related information. It is a statistical term saying how good the particular generated equation is at predicting functional defects. R-Sq. value must be above 85%
- R-Squared adjusted (R-Sq (adj.)) –It is a modification of r-squared used in regression and multiple regression to compare models with different number of explanatory terms. R-Sq. (adj.) must be greater than 85%

Since defects data collected consist of all defects and functional defects while the effort data comprises of effort in test design and effort in all phases prior to system testing, four rounds of regression analysis were performed to cater for these metrics. The kilo lines of code (KLOC) metric were kept as permanent metric in the regression. The results are presented below in Figure 1, Figure 2, Figure 3 and Figure 4:

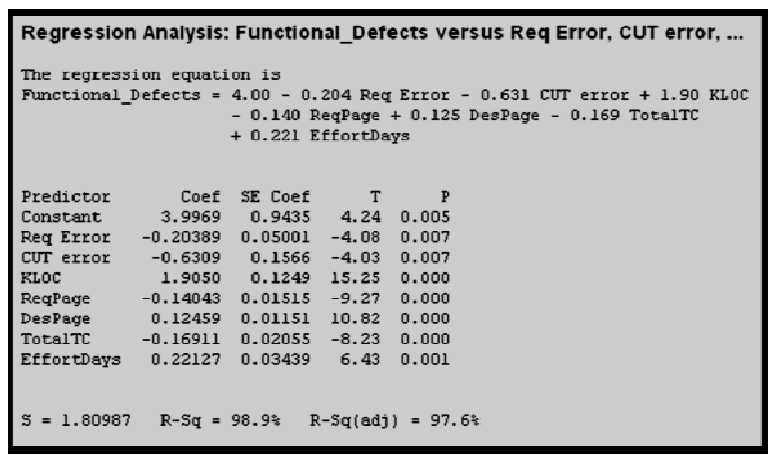

**Figure 1.** Result 1 (Target: Functional Defects; Predictors: Requirement Error, Coding Error, KLOC, Requirement Pages, Design Pages, Total Test Cases, Total Effort Days)

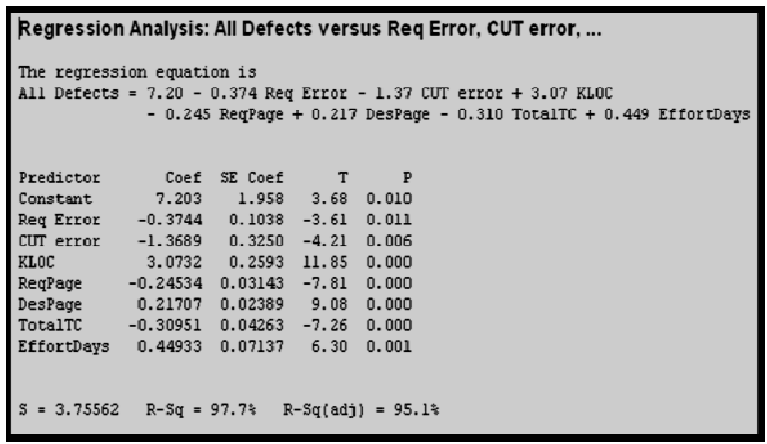

**Figure 2.** Result 2 (Target: All Defects; Predictors: Requirement Error, Coding Error, KLOC, Requirement Pages, Design Pages, Total Test Cases, Total Effort Days)

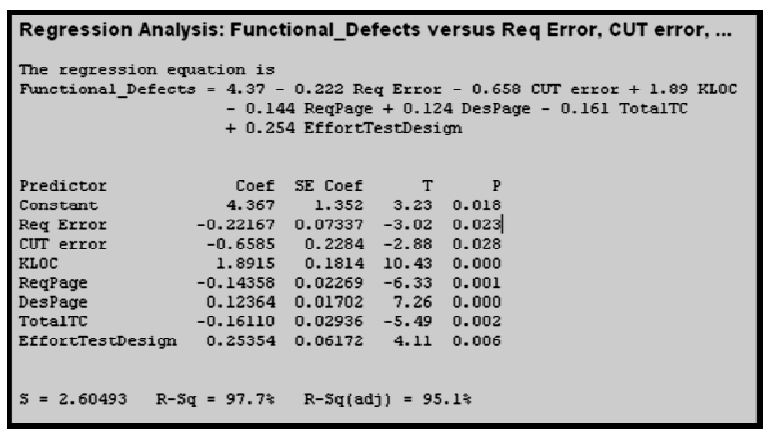

**Figure 3.** Result 3 (Target: Functional Defects; Predictors: Requirement Error, Coding Error, KLOC, Requirement Pages, Design Pages, Total Test Cases, Total Effort Days in Test Design)

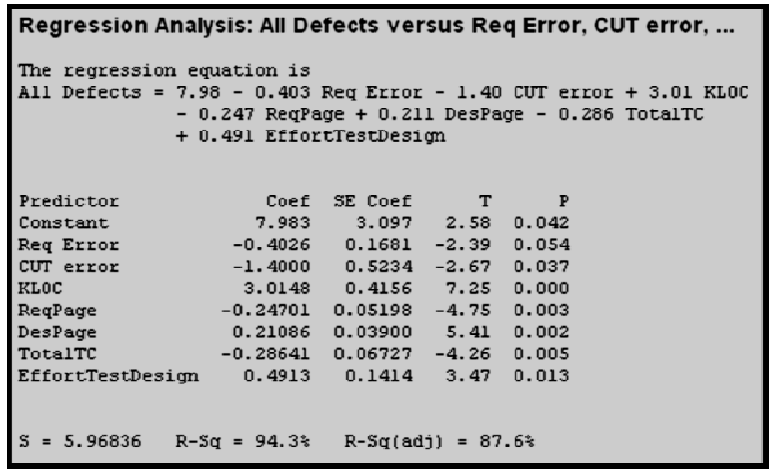

**Figure 4.** Result 4 (Target: All Defects; Predictors: Requirement Error, Coding Error, KLOC, Requirement Pages, Design Pages, Total Test Cases, Total Effort Days in Test Design)

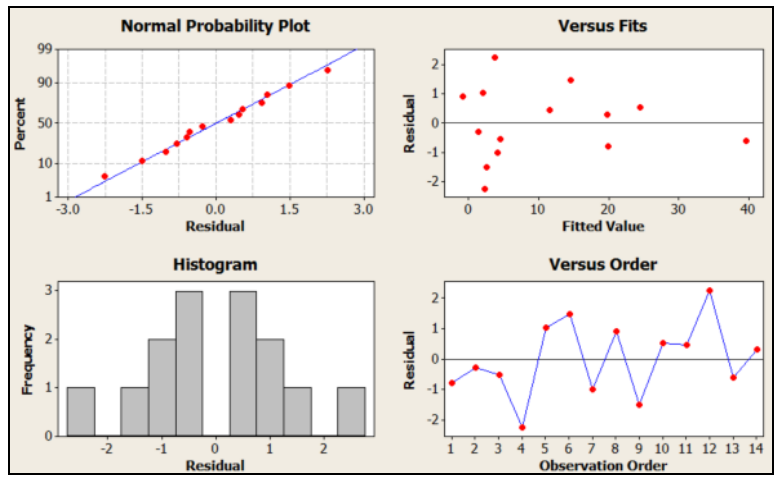

**Figure 5.** Residual plot for Result 1

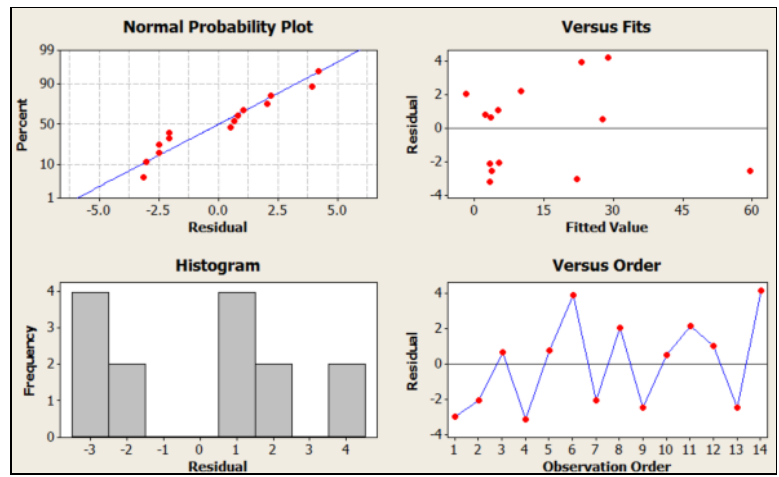

**Figure 6.** Residual plot for Result 2

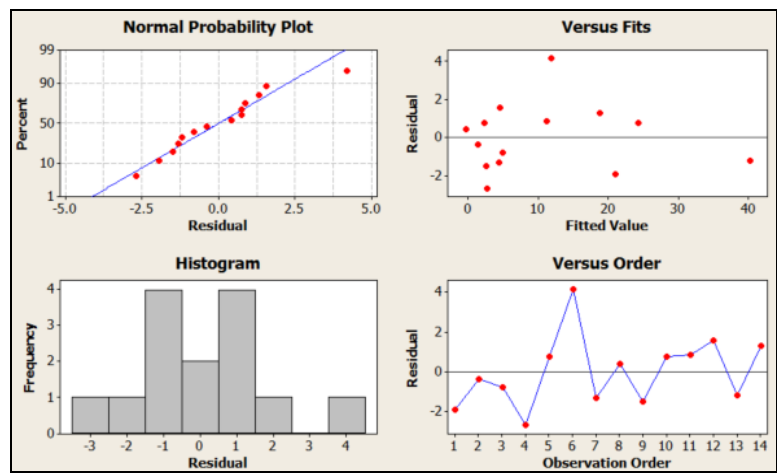

**Figure 7.** Residual plot for Result 3

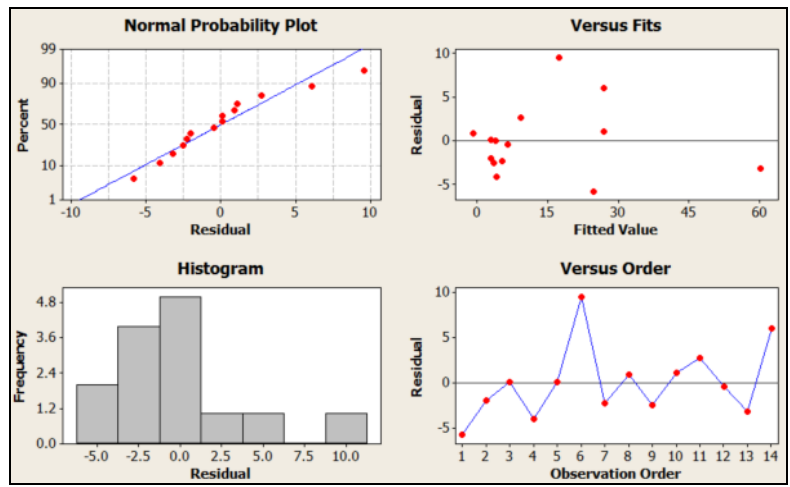

**Figure 8.** Residual plot for Result 4

Based on all four results as represented from Figure 1 to Figure 8, all values of P-value for each predictor are less than 0.05 while all values for R-Sq. and R-Sq. (adj.) are greater than 85%. Therefore, it is difficult to select which regression equation is the right one.

To overcome this, each equation was applied to three new projects which are not part of the data set used to run the regression. This verification activity used Prediction Interval (PI) generated for each new prediction of each new project as the reference to measure the prediction. PI serves as the minimum and maximum range of prediction in which the predicted defects must fall into. Results of the verification for each new project are presented below in Table 3:

**Table 3.** Verification results

| Target | Effort Predictors | Project | Predicted Defects | Actual Defects | 95% PI (min, max) |
|---|---|---|---|---|---|
| Functional Defects | All Tester Effort Prior to System Testing | Project 1 | 182 | 187 | (155, 210) |
| | | Project 2 | 6 | 1 | (0, 14) |
| | | Project 3 | 1 | 1 | (0, 6) |
| All Defects | All Tester Effort Prior to System Testing | Project 1 | 298 | 230 | (241, 356) |
| | | Project 2 | 9 | 9 | (0, 24) |
| | | Project 3 | 2 | 1 | (0, 12) |
| Functional Defects | All Tester Effort in Test Design Prior to System Testing | Project 1 | 183 | 187 | (201, 392) |
| | | Project 2 | 8 | 1 | (0, 19) |
| | | Project 3 | 2 | 1 | (0, 9) |
| All Defects | All Tester Effort in Test Design Prior to System Testing | Project 1 | 296 | 230 | (142, 225) |
| | | Project 2 | 11 | 9 | (0, 37) |
| | | Project 3 | 3 | 1 | (0, 19) |

Figure 9, Figure 10, Figure 11 and Figure 12 show the verification results in graphical format:

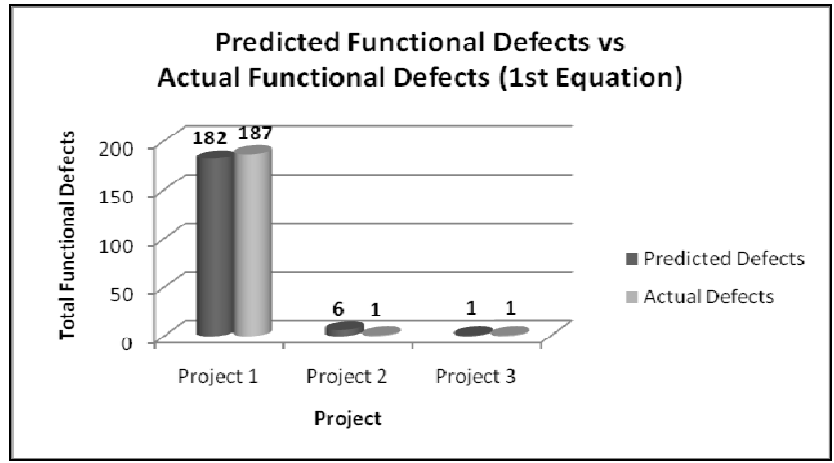

**Figure 9.** Verification result for Equation 1

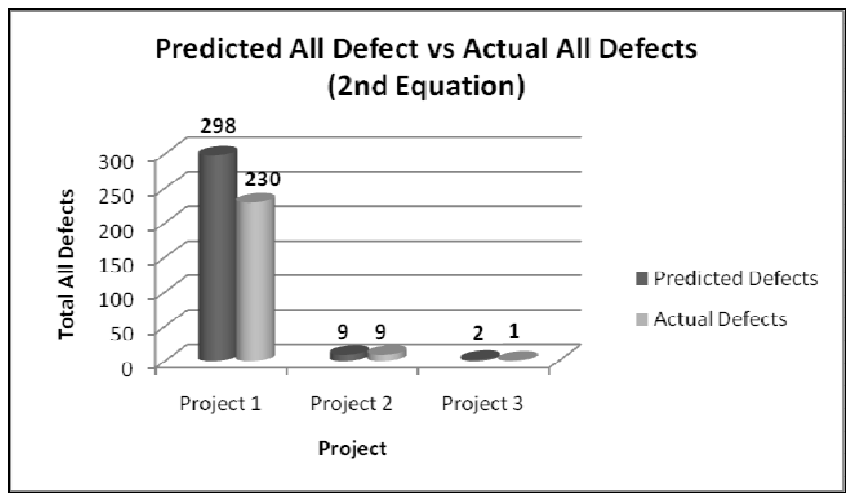

**Figure 10.** Verification result for Equation 2

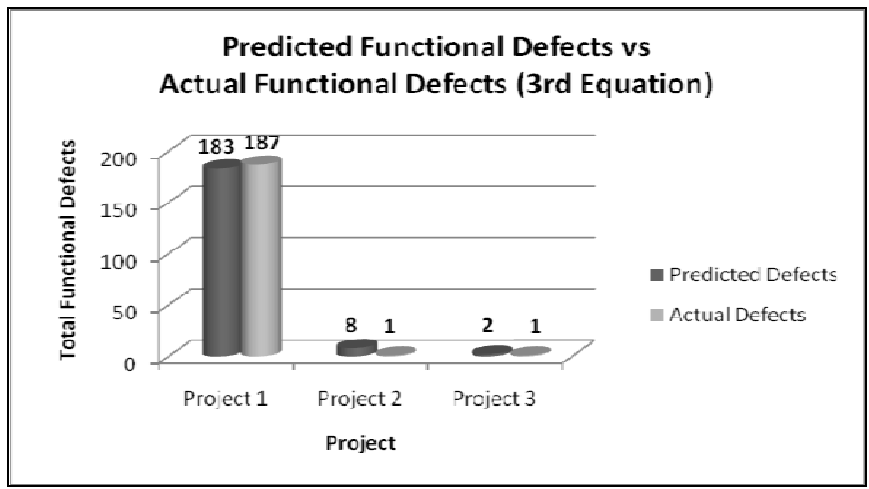

**Figure 11.** Verification result for Equation 3

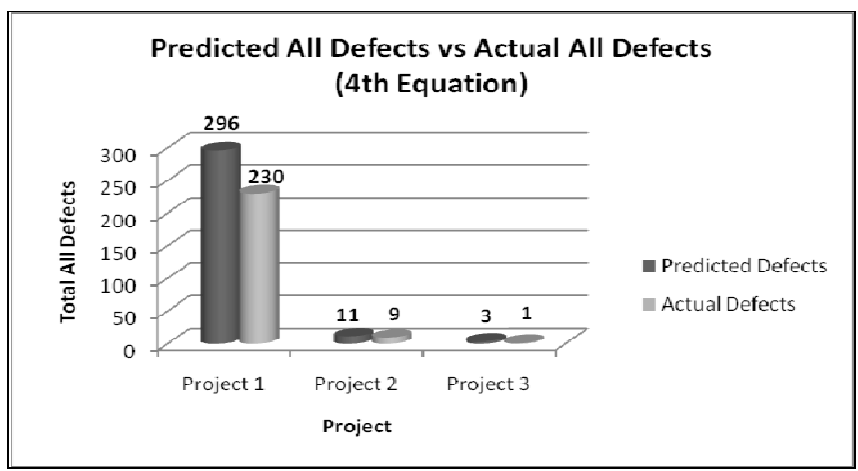

**Figure 12.** Verification result for Equation 4

Since first equation which predicts functional defects and uses total tester efforts in prior phases to testing demonstrated the most promising result, it was the initial defect prediction model for system testing. This is because equation fell between 95% prediction interval and the prediction range is also not as big as others. Thus, the prediction model equation is as below:

Functional Defects = 4.00 - 0.204 Requirement Error - 0.631 Coding Error + 1.90 KLOC –
        0.140 Requirement Page + 0.125 Design Page – 0.169 Total Test Cases +
        0.221Total Effort Days

Based on above equation, it means that the initial defect prediction model for system testing that has been formulated can only predict functional defects by using requirement error, coding error, KLOC, requirement page, design page, total test cases as well as total effort days spent by test engineers as the predictors. This equation still requires future and continuous improvement since it is not sufficient and practical to rely on only one model over long period of time to predict defects due to different software nature and behavior.

## 5. Conclusion

It has been clearly demonstrated that regression analysis has been successfully applied to formulate a prediction model for system testing defects. By using statistical approach such as regression analysis, the research can justify the reasons and significance of metrics from requirement, design and coding phase in predicting defects for system testing. Moreover, it is also explained that in order to have a good model, the prediction must fall between a defined minimum and maximum range so that it is feasible to incorporate and implement defect prediction as part of software development process, particularly test process. Having a prediction of defects using absolute number is not recommended since it requires highly reliable data and rigor data collection to be used for constructing such model.

In carrying out the research, the activities were subjected to several limitations. First, the research only produced one general model due to limited number of data points. Second, data collected is only limited to software development projects in which their metrics are rigorously collected and tracked. Projects that were not involved in metrics collection are no part of the data collected. Third limitation is that this research only focuses on V-shaped development model since that is the process model being adopted by the organization selected for this research. Fourth, data set used in this research is a mix of metrics from web-based and component-based software. Therefore, findings of the research are the final result of using metrics from both software types.

## 6. Future Work

As recommendations for future improvement to the defect prediction model developed in this research, several things could be considered. More variants of the model could be developed by improving the model to predict non-functional test defects such as security testing defects, usability testing defects and performance testing defects. To achieve this, related metrics affecting these non-functional testing must be well defined, collected and tracked. It is also good to have prediction model for different severity of testing defects that are minor, major and critical defects. Other suggestion would be to have more product-centric metrics like number of classes, function points or use case points to be used as the predictors for the model. In supporting the real time process for dynamically generating the latest defect prediction model, a software tool can be developed and used.

It is hoped that the outcome of this research has been able to contribute and expand existing knowledge in software engineering domain, particularly in the area of software testing, software quality management and software process. With the continuous effort in improving such prediction, more high quality software product can be developed in the future.

## References

[1] G. Graham, E.V. Veenendaal, I. Evans, R. Black, "Foundations of Software Testing: ISTQB Certification", Thomson Learning, United Kingdom, 2007.
[2] N.E. Fenton, M. Neil, "A Critique of Software Defect Prediction Models", IEEE Transactions on Software Engineering, vol. 25, no.5, pp.675-689, 1999.
[3] B. Clark, D. Zubrow, "How Good is the Software: A Review of Defect Prediction Techniques", Carnegie Mellon University, USA, 2001.
[4] V. Nayak, D. Naidya, "Defect Estimation Strategies", Patni Computer Systems Limited, Mumbai, 2003.
[5] M. Thangarajan, B. Biswas, "Software Reliability Prediction Model", Tata Elxsi Whitepaper, 2002.
[6] D. Wahyudin, A. Schatten, D. Winkler, A.M. Tjoa, S. Biffl, "Defect Prediction using Combined Product and Project Metrics: A Case Study from the Open Source "Apache" MyFaces Project

Family" In Proceedings of Software Engineering and Advanced Applications (SEAA '08), 34th Euromicro Conference, pp. 207-215, 2008.
[7] I. Sinovcic, L. Hribar, "How to Improve Software Development Process using Mathematical Models for Quality Prediction and Element of Six Sigma Methodology", In Proceedings of the 33rd International Conventionions 2010 (MIPRO 2010), pp. 388-395, 2010.
[8] E.J. Weyuker, T.J. Ostrand, R.M. Bell, "Using Developer Information as a Factor for Fault Prediction", In Proceedings of the Third International Workshop on Predictor Models in Software Engineering (PROMISE'07), pp.8, 2007.
[9] T. Gyimothy, R. Ferenc, I. Siket, "Empirical Validation of Object-Oriented Metrics on Open Source Software for Fault Prediction", IEEE Transactions on Software Engineering, vol. 31, no.10, pp. 897-910, 2005.
[10] J.S. Collofello, "Simulating the System Test Phase of the Software Development Life Cycle", In Proceedings of the 2002 Summer Software Computer Simulation Conference, 2002.
[11] L. Radliński, "Predicting Defect Type in Software Projects", Polish Journal of Environmental Studies, vol.18, no. 3B, pp. 311-315, 2009.
[12] M. Staron, W. Meding, "Defect Inflow Prediction in Large Software Projects", e-Informatica Software Engineering Journal, vol. 4, no. 1, pp. 1-23, 2010.
[13] T. Fehlmann, "Defect Density Prediction with Six Sigma", Presentation in Software Measurement European Forum, 2009.
[14] S.W. Haider, J.W. Cangussu, K.M.L. Cooper, R. Dantu, "Estimation of Defects Based on Defect Decay Model: ED3M", IEEE Transactions on Software Engineering, vol. 34, no. 3, pp. 336-356, 2008.
[15] L. Zawadski, T. Orlova, "Building and Using a Defect Prediction Model", Presentation in Chicago Software Process Improvement Network, 2012.
[16] M. Li, H. Zhang, R. Wu, Z.H. Zhou, "Sample-based Software Defect Prediction with Active and Semi-supervised Learning", Journal of Automated Software Engineering, vol. 19, no. 2, pp. 201-230, 2012.
[17] Z. He, F. Shu, Y. Yang, M. Li, Q. Wang, "An Investigation on the Feasibility of Cross-Project Defect Prediction", Journal of Automated Software Engineering, vol. 19, no. 2, pp. 167-199, 2012.
[18] T.J. Ostrand, E.J. Weyuker, "How to Measure Success of Fault Prediction Models", In Proceedings of Fourth International Workshop on Software Quality Assurance 2007 (SOQUA '07), pp. 25-30, 2007.
[19] L.P. Li, M. Shaw, J. Herbsleb, "Selecting a Defect Prediction Model for Maintenance Resource Planning and Software Insurance", In Proceedings of 5th Workshop on Economics-Driven Software Engineering Research (EDSER '03), pp. 32-37, 2003.
[20] M. D'Ambros, M. Lanza, R. Robbes, "Evaluating Defect Prediction Approaches: A Benchmark and an Extensive Comparison, Journal of Empirical Software Engineering, vol. 17, no. 4-5, pp. 531-577, 2012.

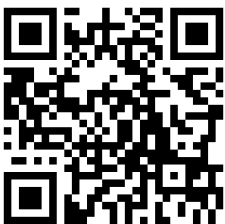

Free download and more information for this paper